\documentclass[preprint]{ptephy_v1}

\preprintnumber{OU-HET-1221}

\usepackage{amsmath,amsfonts}
\usepackage{bm}
\usepackage{physics}
\usepackage{comment}
\usepackage{color}

\begin{document}

\title{A lattice formulation of Weyl fermions on a single curved surface}


\author{Shoto Aoki}
\author{Hidenori Fukaya}
\author{Naoto Kan}
\affil{Department of Physics, Osaka University, Toyonaka 560-0043, Japan}

\begin{abstract}%
  In the standard lattice domain-wall fermion formulation, one needs two flat
  domain-walls where both of the left- and right-handed massless modes appear.
  In this work we investigate a single domain-wall system with a nontrivial curved
  background. Specifically we consider a massive fermion on a $3D$
  square lattice, whose domain-wall is a $2D$ sphere.
  In the free theory, we find that a single Weyl fermion is localized at the wall
  and it feels gravity through the induced spin connection.
  With a topologically nontrivial $U(1)$ link gauge field, however, we 
  find a zero mode with the opposite chirality
  localized at the center where the gauge field is singular.
  In the latter case, the low-energy effective theory is not chiral but vectorlike.
  We discuss how to circumvent this obstacle in formulating lattice chiral gauge theory
 in the single domain-wall fermion system.
\end{abstract}


\maketitle

\section{Introduction}
\label{sec:intro}

Lattice gauge theory has been
successful in understanding and computing nonperturbative dynamics of particle physics. 
Numerical simulation of quantum chromodynamics (QCD) on a flat Euclidean lattice has
been an essential tool for examining the standard model
in various hadronic processes.
Domain-wall fermions, in particular,
defined on a five-dimensional square lattice,
achieve a good control of chiral symmetry,
where the unwanted mixing
between the left- and right-handed quarks
is well suppressed.


In the standard domain-wall fermion formulation \cite{KAPLAN1992342AMethod,Shamir1993Chiral,Furman1995Axial} 
on a flat five-dimensional lattice,
we need at least two domain-walls where the left- and right-handed
massless fermions are localized.
If one of them is decoupled from the gauge fields,
the system may enable formulation of a chiral gauge theory on a lattice 
\cite{Golterman:1992ub,Jansen:1994ym,Kaplan:1995pe, Kikukawa:2000kd}\footnote{
Instead of directly using the domain-wall fermions, 
the effective four-dimensional Dirac operator or known 
as the overlap Dirac operator \cite{Neuberger:1997fp}
was used to formulate some successful examples 
of nonperturbative formulation of lattice chiral gauge theory. 
See Refs.~\cite{Luscher:1998du,Luscher:1999un,Kikukawa:2000kd} and the references therein.
}.
In order to decouple the unwanted modes, 
a smooth elimination of gauge fields
has been considered in \cite{
Grabowska:2015qpk,Grabowska:2016bis} 
but it turned out that this task is not that easy.
For instance, the cobordism invariance of the instanton
background forces the two domain-walls to have the same number
of zero modes \cite{Okumura:2016dsr} (see also related works \cite{Hamada:2017tny, Makino:2016auf,Makino:2017pbq, Ago:2019yzj}).
This indicates that the low energy theory
has both of the left- and right-handed modes.
Another approach is the so-called symmetric mass generation \cite{
Wen:2013ppa,Wang:2013yta,Kikukawa:2017ngf,Kikukawa:2017gvk,Wang:2018jkc,you2018symmetric,Tong:2021phe,Zeng:2022grc,Wang:2022ucy,Lu:2023emm,Guo:2023rnz,Lu:2023cev,Liu:2023msa,Golterman:2023zqf},
which gaps out the unwanted chiral modes by generating mass without breaking the chiral symmetry.
This, however, requires a highly nonperturbative dynamics.

In our recent works \cite{Aoki:2022cwg,Aoki:2022aez} 
we have been studying a massive fermion system with a nonflat single domain-wall.
In the case with an $S^1$ domain-wall embedded into a flat $2D$
square lattice, we observed that massless Dirac fermion appears
on the wall, having a definite eigenvalue of the
gamma matrix perpendicular to the domain-wall.
Similarly, we observed that a two-flavor massive fermion,
which respects the time-reversal symmetry,
yields a massless Dirac fermion on the $S^2$ domain-wall embedded in flat three dimensions.
Also in this case, the edge-localized modes are the
eigenstates of the gamma matrix accompanied 
by an operator exchanging the flavor index.

Moreover, it was found through the Dirac operator spectrum that
the edge-localized modes feel gravity
via the induced spin connection
on the curved domain-wall.
This is a realization of the Einstein equivalence principle,
which tells that any acceleration is indistinguishable from gravity.
We have numerically confirmed that the rotational symmetry
is monotonically recovered toward the classical continuum limit
of the higher-dimensional square lattice.
The limit of the Dirac operator eigenvalue spectrum
is consistent with that in the continuum theory, which has
a manifest gauge symmetry under the
general coordinate transformation.

In this work, we extend our study on the curved domain-wall fermions
in two new directions.
One is to employ Shamir's formulation \cite{Shamir1993Chiral,Furman1995Axial}.
We consider a $2D$ spherical domain-wall embedded
into a $3D$ square lattice,
but we ignore the lattice points outside of the domain-wall,
which is equivalent to taking the fermion mass to $+\infty$.
The second is to consider a single-flavor fermion,
which loses the time-reversal symmetry in three dimensions.
Then the edge-localized modes should represent not Dirac but Weyl fermions.
From the Dirac operator spectrum, we investigate the gravity
induced on the wall, and examine its continuum limit.
A preliminary result of this work was presented in Ref.~\cite{Aoki:2024pad}.

In this setup, a single massless fermion with a definite chirality
is localized on a single domain-wall.
Namely, it looks as if a Weyl fermion on
an $S^2$ surface is nonperturbatively regularized on a lattice.
In recent publications \cite{Kaplan:2023pxd,Kaplan:2023pvd}
this single-domain-wall system was proposed
as a lattice regularization of chiral gauge theories.

Contrary to the expectation, when it is coupled to the $U(1)$ gauge field,
we observe a nontrivial dynamics where the low-energy theory
becomes not chiral but rather vectorlike.
This happens when the gauge field on the $S^2$ has a nontrivial topological charge.
The Gauss law in three dimensions requires a singular source
of the gauge flux inside the spherical domain-wall.
This configuration is an analog of a magnetic monopole in three spatial dimensions,
whose magnetic flux penetrates the $S^2$ surface.
The additive mass renormalization through the Wilson term
due to the singular gauge field becomes strong enough
to flip the sign of the effective mass of the fermion,
and another domain-wall is created around the singularity.
The dynamically created domain-wall captures a single fermion zero mode
with the opposite chirality to the original edge modes on the $S^2$.
This dynamical creation of domain-walls was already reported
in our previous work on the Dirac fermion system \cite{Aoki:2023lqp,Kan:2023latticepro}
which gives a microscopic description of the Witten effect 
\cite{Witten:1979Dyons}: 
the magnetic monopole gains an electric charge because
the electron zero mode is bound to
the small but finite spherical domain-wall around the monopole.
We discuss how to circumvent this obstacle in formulating chiral gauge theory.

The rest of this paper is organized as follows.
In Sect.~\ref{sec:ShamirDW} we show that the edge-localized modes generally exist and
they feel the induced gravity.
In Sect.~\ref{sec:LatticeFree} we analyze the lattice 
domain-wall fermion system numerically.
Our result shows that a Weyl fermion appears on the $S^2$ domain-wall,
as is expected.
In Sect.~\ref{sec:LatticeU1}, however, we show that 
an oppositely chiral zero mode appears when 
the gauge field has a point-like singularity.
We discuss in Sect.~\ref{sec:discussion} how to avoid the
unwanted zero modes with the opposite chirality.
A summary is given in Sect.~\ref{sec:summary}.

\section{Continuum analysis of $3D$ dimensional Shamir domain-wall fermion with a single $S^2$ boundary}
\label{sec:ShamirDW}

We first consider a Dirac fermion action in a $3D$
flat continuum spacetime $\mathbb{R}^3$,
\begin{equation}
S = \int d^3 x\; \bar{\psi}D\psi(x),
\end{equation}
where the Dirac operator with a position-dependent mass $m(x)$ is given by
\begin{equation}
  \label{eq:D}
  D= \sum_{i=1}^{3} \sigma^i \qty(\pdv{x^i} -iA_i(x))+ m(x).
\end{equation}
Here we have introduced a background $U(1)$ gauge field $A_i(x)$
and $\sigma^i$ denote the Pauli matrices representing the gamma matrices in three dimensions.

We assume that the mass function $m(x)$ has a spherical domain-wall structure:
\begin{align}
\label{eq:massterm}
  m(x)=\left\{
  \begin{array}{cc}
  -m & \text{for $|x|\le r_0$}\\
  +M & \text{otherwise}
  \end{array}\right.,
\end{align}
where we assume $m>0, M>0$.
In order to obtain the Shamir domain-wall fermion Dirac operator,
we take the $M\to +\infty $ limit, which is equivalent to
imposing a boundary condition
\begin{equation}
\sigma_r \psi(x) = +\psi(x)\;\;\;\text{at $|x|=r_0$}.
\end{equation}
Here the chirality operator $\sigma_r$ is given by
\begin{equation}
\sigma_r = \frac{\sum_i \sigma^i x_i}{|x|},
\end{equation}
which is well-defined at any point on the $S^2$ surface $|x|=r_0$.
In the same way, the conjugate operator $D^\dagger$ requires
the boundary condition with the opposite chirality $\sigma_r=-1$.

We show below that the chiral edge-localized modes generally exist on 
the curved spherical domain-wall,
at least approximately in Sect.~\ref{sec:general-edge}.
We also derive in Sect.~\ref{sec:exact} an exact eigenfunction 
when the gauge field configuration respects the rotational symmetry.

\subsection{Edge-localized modes in general background gauge field}
\label{sec:general-edge}

First, let us confirm the existence of the edge-localized modes
in a general $U(1)$ gauge field background.
Here we assume that the gauge field is smooth everywhere
without any singularity.

In the polar coordinates and taking the radial gauge condition $A_r=0$, 
the Dirac operator inside the domain-wall $|x|\leq r_0$ is expressed by
\begin{align}
    D
    =& \sigma_r \qty( \pdv{r} + \frac{1}{r} - \frac{r_0}{r}iD^{S^2}) - m,
\end{align}
where the Hermitian operator $iD^{S^2}$ is defined by
\begin{equation}
  iD^{S^2} := \frac{1}{r_0}\left(1-i\sigma_2\exp(i\phi \sigma_3)\left({\partial \over \partial \theta}-i\hat{A}_\theta\right)-{1 \over \sin\theta}\sigma_r\sigma_2\exp(i\phi\sigma_3)\left({\partial \over \partial \phi}-i \hat{A}_\phi\right)\right).
\end{equation}
Here we have introduced the dimensionless vector potentials: 
$\hat{A}_\theta:=r A_\theta$ and $\hat{A}_\phi:=r \sin\theta A_\phi$.

In fact, we can identify $D^{S^2}$ as the effective Dirac operator
on $S^2$ for the low-energy edge-localized modes.
Taking a gauge transformation or the Euclidean version of the local Lorentz transformation,
\begin{align}
    R(\theta,\phi) =\exp( i \theta \sigma^2/2 ) \exp( i \phi \sigma^3/2 ),
\end{align}
as well as a $U(1)$ gauge transformation, we obtain the $2D$ massless Dirac operator:
\begin{align}
\label{eq:effectiveD}
    e^{-i \frac{\phi}{2}} R(\theta,\phi) iD^{S^2} e^{i \frac{\phi}{2}} R(\theta,\phi)^{-1}
    = -\frac{1}{r_0}\sigma_3 \left[\sigma_1 \left(\pdv{}{\theta}-i\hat{A}_\theta\right) 
+\frac{\sigma_2}{\sin \theta} \qty( \pdv{}{\phi}- i \hat{A}_\phi+\frac{i}{2 } -\frac{i\cos\theta}{2} \sigma_3 )\right].
\end{align}
We can see that the spin connection $\sigma_3\cos \theta/2$ is induced\footnote{
To be precise, the induced connection is spin-c with an
additional $U(1)$ part $\frac{i}{2}$, 
which, however, can be eliminated by the 
gauge transformation. We keep this unphysical $U(1)$ part nonzero 
to make the analysis simpler, since otherwise, the gauge transformation 
requires multiple-patch covering of $\mathbb{R}^3$
to avoid multivaluedness of the gauge transformation with respect to $\theta$.}.
It is also important to note that the rotated $2D$ Dirac operator 
anticommutes with the chirality operator 
$R(\theta,\phi)\sigma_rR(\theta,\phi)^{-1}=\sigma_3$.

Let $\chi(\theta, \phi)$ be a function in the domain of $iD^{S^2}$ at $r=r_0$.
Then we can construct an edge-localized mode
\begin{align}
 \psi_+^e = \frac{1}{r}\exp[-m(r_0-r)]P_+\chi(\theta, \phi),
\end{align}
in relation to which the Dirac operator effectively acts as
\begin{align}
 D = P_- (iD^{S^2}) P_+, 
\end{align}
with the chiral projection operators $P_\pm :=(1\pm \sigma_r)/2$.
Here we have used an approximation in the large $m$ limit,
\begin{align}
 \frac{1}{r}iD^{S^2}\exp[-m(r_0-r)]\sim \frac{1}{r_0}iD^{S^2}\exp[-m(r_0-r)].
\end{align}
Thus, the edge-localized mode $\psi_+^e$ represents one massless fermion 
with positive chirality on the single $S^2$ surface with radius $r_0$.

When we take $M$ in Eq.~(\ref{eq:massterm}) to negative infinity keeping the sign of $m$ unchanged,
the boundary condition requires the negative chirality at $r=r_0$,
which prohibits the edge-localized modes.
If we take both $m$ and $M$ negative, the negatively chiral edge-mode 
\begin{align}
 \psi_-^e = \frac{1}{r}\exp[m(r_0-r)]P_-\chi(\theta, \phi),
\end{align}
appears. 
The case with $mM<0$ is analogous to a normal insulator
where no edge-localized mode appears,
and $mM>0$ corresponds to a topological insulator where 
the existence of the edge excitation is topologically protected.
Thus, our curved domain-wall fermion system with $mM>0$
has the edge-localized modes, which
describe a single Weyl fermion on a single curved surface.

The edge-localized Weyl fermion feels inertial force resulting from the constrained motion. As explicitly shown in Eq.~(\ref{eq:effectiveD}), this force can be identified as gravity through the
induced spin connection. 
This is consistent with Einstein's equivalence principle,
which indicates that any constraint force can be identified as gravity.
Since the edge modes are constrained on the curved $S^2$ surface,
gravity is naturally induced on them.
Interestingly, the induced gravity reflects only intrinsic geometry
and the extrinsic information is invisible.
This is not special for $S^2$ but true on general curved domain-walls
as discussed in Ref.~\cite{Aoki:PhD2024}.
The extrinsic curvature contribution is 
suppressed exponentially in the large-$m$ limit.

\subsection{Exact edge-localized modes with rotational symmetry}
\label{sec:exact}

Next we exactly solve the eigenproblem of the
curved domain-wall fermion Dirac operator
in the case where the system has a rotational symmetry.
Let us take the radial gauge $A_r=0$ and assume
that the field strength $F_{\mu\nu}$ perpendicular to the
domain-wall is uniformly distributed.
Such $U(1)$ backgrounds are classified by 
the topological charge,
\begin{align}
 n &= \frac{1}{4\pi}\int_{S^2}d^2x \epsilon^{\mu\nu}F_{\mu\nu},\;\;\;F_{12}=-F_{21}=\frac{n}{2r_0^2}.
\end{align}

Our target gauge field background in the whole $3D$ space
is achieved by putting a ``monopole'' at the origin\footnote{
It may not be adequate to name this configuration a ``monopole'' since our gauge potential
is considered not in $3D$ space but in $2+1$-dimensional spacetime.
But the computation is completely parallel to the case with the monopole 
in three spatial dimensions we discussed in Ref.~\cite{Aoki:2023lqp}. 
}.
The vector potential generated by the monopole is 
\begin{align}
 \label{eq:Amu}
    A_1=\frac{-q_m y}{r(r+z)},~A_2=\frac{q_m x}{r(r+z)},~A_3=0,
\end{align} 
where $q_m$ is the magnetic charge of the monopole,
which satisfies the Dirac's quantization condition $q_m=\frac{n}{2}~(n\in\mathbb{Z})$.
To be precise, this expression has a singularity at $x=y=0$ and $z<0$, which is the so-called Dirac string. 
In order to eliminate the Dirac string, we need another set of 
the gauge potentials in the $z\le 0$ region but here we simply ignore the
neighborhood of the string. 
In fact, the Dirac string automatically disappears in the 
lattice regularization that we will discuss in the next section.
The field strength we have obtained is
\begin{align}
    F_{ij}= \partial_i A_j - \partial_j A_i=q_m \epsilon_{ijk} \frac{x_k}{r^3}.
\end{align}

With the above spherically symmetric gauge field background,
the effective $2D$ Dirac operator $iD^{S^2}$ can be written as
\begin{equation}
\label{eq:2DD}
  D^{S^2} = \frac{1}{r_0}\left[\sigma^i \left(L_i+n\frac{x_i}{2r}\right) +1\right],
\end{equation}
where $L_i$ is the orbital angular momentum operator in the presence of the monopole,
\begin{equation}
 L_i = -i\epsilon_{ijk}x^j\left(\frac{\partial}{\partial x^k}-iA_k\right)-n\frac{x_i}{2r}.
\end{equation}

Instead of $D$ itself, let us solve the eigenproblem of the Hermitian operator $D^\dagger D$.
Note that $D^\dagger D$ commutes with the total angular momentum operators
$J^i = L^i +\sigma^i/2$ as well as $D^{S^2}$ so that it is convenient
to consider the simultaneous eigenstates of $\bm{J}^2=\sum_i(J^i)^2$, $J^3$ and $D^{S^2}$. 
Let $j$ be the highest weight of $\bm{J}$, which takes discrete values\footnote{
    The lower bound of $j$ comes from the nontrivial contribution from
    monopole, which is proportional to $n$.
    }
\begin{align}
j= \frac{ \abs{n} -1 }{2},~ \frac{ \abs{n} -1 }{2}+1, \frac{ \abs{n} -1 }{2}+2\cdots,
\end{align}
with a constraint $j\ge 0$. 
The eigenfunction denoted by $\chi_{j,j_3, \pm}$ satisfies
\begin{align}
    \bm{J}^2 \chi_{j,j_3, \pm} &=j(j+1) \chi_{j,j_3, \pm}, \\
    J_3 \chi_{j,j_3, \pm} &= j_3 \chi_{j,j_3, \pm}, \\
\label{eq:Ds2}
     iD^{S^2}\chi_{j,j_3, \pm}&= \pm \frac{\nu}{r_0}\chi_{j,j_3, \pm},
\;\;\;\nu=\sqrt{ \qty(j+\frac{1}{2})^2-\frac{n^2}{4}  }, \\
    \sigma_r \chi_{j,j_3, \pm} &= \chi_{j,j_3, \mp}.
\end{align}
Here we have used facts that $(D^{S^2})^2$ is written as a linear combination of
$\bm{J}^2$, $\sigma_r$ and identity, and that $D^{S^2}$ anticommutes
with $\sigma_r$. 
As will be shown below, the edge-localized eigenvalue of $D^\dagger D$ depends only on $j$.
Therefore the degeneracy of the eigenvalue is $2(2j+1)$ \cite{Wu:1976ge,Aoki:2023lqp}.

Since the chirality operator $\sigma_r$, which determines the boundary condition at $r=r_0$, 
anticommutes with $D^{S^2}$,
the eigenstate $\psi$ of $D^\dagger D$ is written by a linear combination of $\chi_{j,j_3 ,\pm}$:
\begin{align}
    \psi= \frac{1}{\sqrt{r}} \left[g_+(r) \chi_{j,j_3 ,+} + g_-(r) \chi_{j,j_3 ,-} \right].
\end{align}
Substituting this into the equation $D^\dagger D \psi = E^2\psi$, where we assume $m^2>E^2$, 
we find that the functions $g_\pm(r)$ satisfy the modified Bessel equations
with a rescaled variable $z=\kappa r$ where $\kappa=\sqrt{m^2-E^2}$,
\begin{align}
\label{eq:Bess}
 \left[\frac{\partial^2}{\partial z^2}+\frac{1}{z}\frac{\partial}{\partial z}-\frac{(\nu\mp 1/2)^2}{z^2}-1\right]g_\pm =0.
\end{align}
From the regularity condition at $r=0$, we find the solutions in terms of 
the modified Bessel function of the first kind:
 $g_+(r)=AI_{\nu-1/2}(\kappa r)$ and $g_-(r)=BI_{\nu+1/2}(\kappa r)$
with numerical constants $A$ and $B$, respectively.
Since $I_\mu(z)$ are an exponentially increasing function of $z$,
the solutions are exponentially localized at the edge, $r=r_0$.

The boundary condition at $r=r_0$: $\sigma_r\psi=+\psi$ and $\sigma_r D\psi=-D\psi$ 
leads to two equations
\begin{align}
\label{eq:bceq}
 AI_{\nu-1/2}(\kappa r_0) &= BI_{\nu+1/2}(\kappa r_0),\\
 (\kappa B-mA)I_{\nu-1/2}(\kappa r_0) &=-(\kappa A-mB)I_{\nu+1/2}(\kappa r_0),
\end{align}
which determine the coefficients $A,B$ up to normalization,
and the eigenvalue $E^2$.
According to the asymptotic form $I_\mu(z)\sim \frac{e^z}{ \sqrt{2\pi z} } \qty( 1- \frac{4\mu^2-1}{8z}  )$, the eigenvalue converges to
\begin{equation}
 \label{eq:evalcont}
 E^2={\nu^2 \over r_0^2}
\end{equation}
in the large-$\kappa r_0$ limit, which is independent of the mass parameter $m$.
Moreover, the eigenvalue is shared with that of the massless
$2D$ Dirac operator $(iD^{S^2})^2$ shown in Eq.~(\ref{eq:Ds2}).
It is important to remark in the free case with $n=0$
that the eigenvalue is gaped from zero, which
is a gravitational effect \cite{aoki2021chiral,Aoki:2022cwg}.

The case of $j=\frac{\abs{n}-1}{2}$ with $n\neq 0$, 
is qualitatively different from the above result.
Let $\chi_{j,j_3,0}$ be an eigenstate satisfying
\begin{align}
    \bm{J}^2 \chi_{j,j_3, 0} &=j(j+1) \chi_{j,j_3, 0} \\
    J_3 \chi_{j,j_3, 0} &= j_3 \chi_{j,j_3, 0} \\
    iD^{S^2}\chi_{j,j_3, 0}&= 0\\
    \sigma_r \chi_{j,j_3, 0} &= \text{sign}(n) \chi_{j,j_3, 0}.
\end{align} 
Assuming the edge-localized mode has the following form,
\begin{align}
    \psi_0= \frac{1}{\sqrt{r}} f(r) \chi_{j,j_3,0},
\end{align}
we find that $f(r)$ satisfies the same equation as Eq.~(\ref{eq:Bess}), but with $\nu=0$ 
and thus we obtain
\begin{align}
    f(r)=CI_{1/2}(\kappa r)=C\sqrt{\frac{2}{\pi \kappa r}}\sinh(\kappa r),
\end{align}
with a normalization constant $C$.

It is interesting to note that the chiral 
boundary condition $\sigma_r\psi_0(r=r_0)=+\psi_0(r=r_0)$
is never satisfied with $n<0$.
Also, another boundary condition $\sigma_r D\psi_0(r=r_0)=-D\psi_0(r=r_0)$
is only satisfied by $D\psi_0(r=r_0)=0$ when we take 
the $m\to \infty$ limit.
This reflects the Atiyah-Singer (AS) index theorem \cite{Atiyah1963TheIndexOfEllipticOperator} on the sphere $S^2$,
\begin{equation}
 {\rm Ind}D^{S^2} = \frac{1}{4\pi}\int_{S^2}d^2x \epsilon^{\mu\nu}F_{\mu\nu} = n.
\end{equation}
The Weyl fermion with positive chirality cannot 
give a negative contribution to the index when $n<0$.
Only in the $m\to \infty$ limit do we
have an exactly chiral zero mode of both 
$D^\dagger D$ and $iD^{S^2}$.

Thus, we have obtained the exact massless edge-localized modes
with the positive chirality, describing a
Weyl fermion on the sphere with radius $r_0$.
Their existence is guaranteed even in the 
presence of nontrivial gauge field background.
This single domain-wall fermion system 
looks a good candidate for nonperturbatively formulating
chiral gauge theory on a lattice.
Note, however, that the chirality is not perfect 
unless we take the $mr_0\to \infty$ limit.
In particular, it is not obvious in the above analysis 
what will happen to the chiral zero mode
when $mr_0$ is finite.
So far we have ignored the short-distance 
singularity at the location of the monopole.
As will be seen below, a highly nonperturbative 
and careful short-distance analysis 
is needed to understand this problem.

\section{Lattice analysis in free theory}
\label{sec:LatticeFree}

In this section we discuss a lattice regularization
of the curved domain-wall fermion system.
In the free fermion theory with trivial link variables
we show that a single Weyl fermion appears on
a single spherical domain-wall.
We also discuss its continuum and finite volume systematics.

For discretization of the Dirac operator,
we employ the Wilson Dirac operator
\begin{align}
    D_W= &\sum_{i=1}^{3} \sigma^i \frac{ \nabla_i - \nabla_i^\dagger }{2a} + \sum \frac{w}{2a} \nabla_i \nabla_i^\dagger -m,
\end{align}
where the lattice spacing is denoted by $a$, and we assume that $m>0$.
The difference operator in the $i$-direction for a spinor field located at the lattice site $x$
is defined by
\begin{align}
(\nabla_i \psi )_x= \psi_{x+\hat{i}} -\psi_x, 
\end{align}
with the unit vector $\hat{i}$ in the corresponding direction.
As the standard choice, we take the coefficient of the Wilson term $w=1$.
With this choice, the Dirac operator can be written as
\begin{align}
 D_W = -\frac{1}{a}\sum_{i=1}^{3} \left(P^i_-\nabla_i +P^i_+\nabla_i^\dagger\right)-m,\;\;\;
P_\pm^i =\frac{1\pm \sigma_i}{2}.
\end{align}

\begin{figure}
    \centering
    \includegraphics[scale=0.7,bb=0 0 354 351]{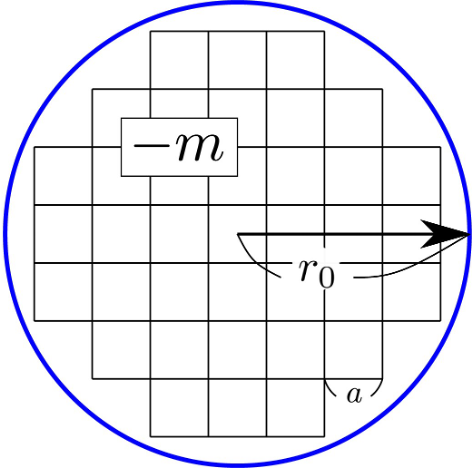}
    \caption{A $2D$ slice at $z=1/2$ of the spherical 
domain-wall in our lattice space. The hopping
to the outside of the sphere $r=r_0$ is cut off. Adapted from \cite{Aoki:2024pad}.}
    \label{fig:CDW}
\end{figure}

It was shown by Shamir \cite{Shamir1993Chiral} 
in the standard flat domain-wall fermion that
there is no need to impose a chiral boundary condition
on the lattice but it is enough to simply to set
$\psi_x=0$ outside the domain-wall.
The same is true for our curved case. 
As shown in Fig.~\ref{fig:CDW} where a $2D$ slice at $z=1/2$ is drawn, 
we impose 
\begin{equation}
 \psi_x=0 \;\;\;\mbox{if $|x-x_0|\ge r_0$},
\end{equation}
where the center of the domain-wall is put at the origin $x_0=(0,0,0)$.
Note that our lattice sites are put 
at half-integral points: $x=((2n_1-1)/2,(2n_2-1)/2,(2n_3-1)/2)$
with integer values $n_i$.
Put $x$ be one of the inner nearest-neighbor sites of the domain-wall, where
$x+\hat{j}$ is put outside of the wall.
Then the Dirac operator acts as 
\begin{align}
 (D_W\psi)_x = -\frac{1}{a}\left[P^j_-\left(-\psi_x\right) -P^j_+\left(\psi_x-\psi_{x-\hat{j}}\right)\right]+\cdots
\end{align}
which imposes the boundary condition $P^j_-\psi_x=0$ in the continuum limit $a\to 0$,
while $P^j_+\psi_x$ can be taken arbitrarily, as far as $\psi_x-\psi_{x-\hat{j}} \sim O(a)$.
With this simple boundary condition, the Hermitian conjugate of the 
Wilson Dirac operator $D^\dagger_W$ is canonically defined.

To be precise, our domain-wall is not a smooth sphere but
a rather digitized zig-zag surface like a block toy
as depicted in Fig.~\ref{fig:CDW}.
It is, therefore, important to check if the rotational
symmetry is recovered in the continuum limit.
In our previous work on Dirac fermions \cite{aoki2021chiral}, 
we already obtained positive numerical evidence.
The rotational symmetry of the lowest eigenmodes
was almost linearly recovered in the $a\to 0$ limit.
We will address this issue for the Weyl fermion at the end of this section.

We numerically solve the eigenproblem of $D_W^\dagger D_W$
as well as $D_W D_W^\dagger$ which are both Hermitian.
With the obtained normalized eigenfunction $\phi_k(x)$, 
where $k$ is the label for the low-lying modes,
we measure the expectation value
\begin{align}
 \sum_x\phi_k^\dagger(x)\sigma_r(x)\phi_k(x),
\end{align}
where the chirality operator $\sigma_r(x)=\sum_i\sigma_i x^i/|x|$ 
is unambiguously defined at all our half-integral lattice sites.
We also scan the local chirality, defined by
\begin{align}
 \label{eq:local chirality}
 \phi_k^\dagger(x)\sigma_r(x)\phi_k(x)/\phi_k^\dagger(x)\phi_k(x),
\end{align}
for each lattice site $x$ in order to 
study its spatial distribution.
This turns out to be useful in analyzing
the case with a nontrivial $U(1)$ gauge field background in the next section.

\begin{figure}
\centering
\includegraphics[bb=0 0 576 346,width=\textwidth]{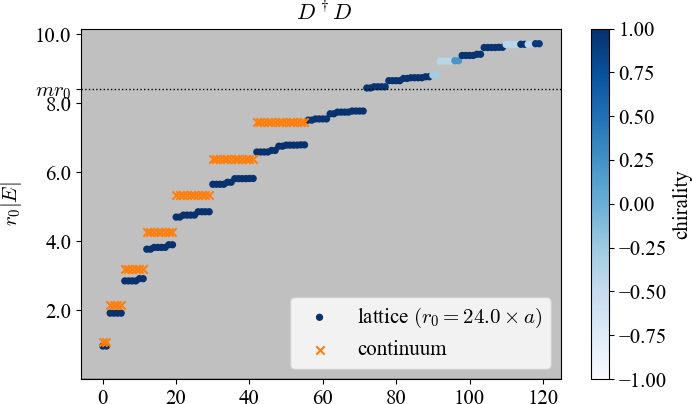}
\caption{The spectrum of $D_W^\dagger D_W$ at $r_0=24 a$ and $ma=0.35$. 
The square-rooted value $r_0|E|$ is plotted.
The color gradation shows the chirality expectation value.
The dotted horizontal line indicates the value of mass $mr_0$, 
above which the eigenmodes are extended to the bulk. 
Our low-lying eigenvalues below $mr_0$ of the edge-localized modes 
agree well with the continuum prediction marked by cross symbols.}
\label{fig:DdaggerDeigenvalue_r0=8.0_n=0_m=14perL}
\end{figure}

In Fig.~\ref{fig:DdaggerDeigenvalue_r0=8.0_n=0_m=14perL},
we plot the eigenvalue spectrum of $D_W^\dagger D_W$ at $r_0=24a$ and $ma=0.35$.
Solving $D_W^\dagger D_W \phi_k =E_k^2\phi_k$, its
square root normalized by $r_0$: $r_0|E|$ is plotted by filled circle symbols.
The color gradation shows the chirality expectation value.
We can see that the low-lying modes below $mr_0$, 
shown by the dotted horizontal line, 
have the positive chirality.
The obtained eigenvalues agree with the continuum prediction marked by cross symbols 
in which the finite volume correction 
to the infinite volume value $r_0|E|=\nu=1,2,3,\cdots$ in Eq.~(\ref{eq:evalcont})
is taken into account.
The $2(j+1/2)$ (or $2\nu$ in this case) degeneracy with $j=1/2,3/2,\cdots$ looks also consistent.
The gravitational effect through the induced spin connection,
or the gap from zero is clearly seen.

\begin{figure}
    \centering
    \includegraphics[bb=0 0 576 346,width=\textwidth]{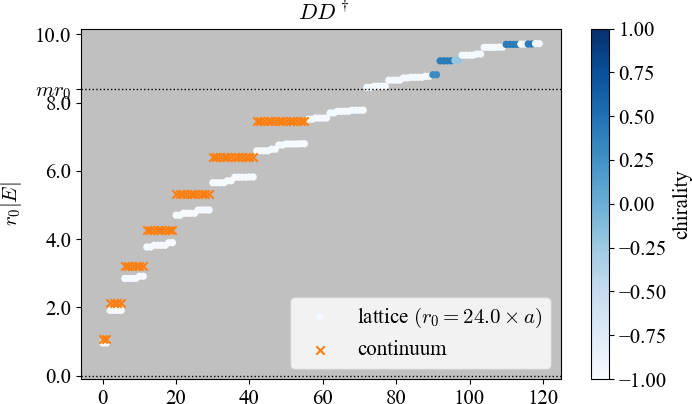}
    \caption{The spectrum of $D_WD_W^\dagger$. 
The lattice setup is the same as Fig.~\ref{fig:DdaggerDeigenvalue_r0=8.0_n=0_m=14perL}.}
    \label{fig:eigenvalue_r0=8.0_n=0_m=14perL}
\end{figure}

We also plot the eigenvalue spectrum of $D_WD_W^\dagger$ on the same lattice
at Fig.~\ref{fig:eigenvalue_r0=8.0_n=0_m=14perL}.
The spectrum is the same as that of $D_W^\dagger D_W$ but the chirality is opposite.

\begin{figure}
    \begin{minipage}[b]{0.48\linewidth}
    \centering
    \includegraphics[bb= 0 0 496 286,width=\textwidth]{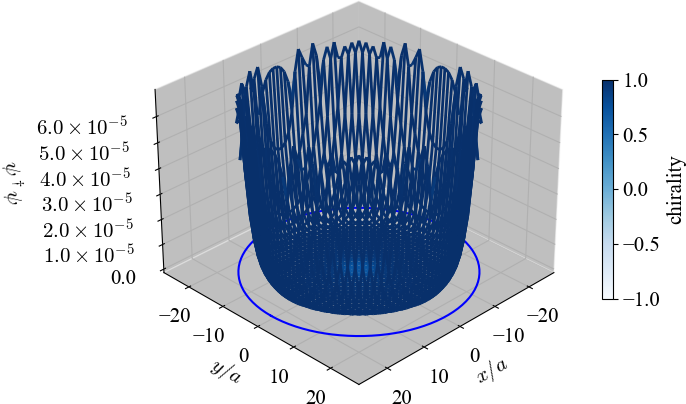}
    \end{minipage}
    \hfill
    \begin{minipage}[b]{0.48\linewidth}
        \centering
        \includegraphics[bb=0 0 496 286,width=\textwidth]{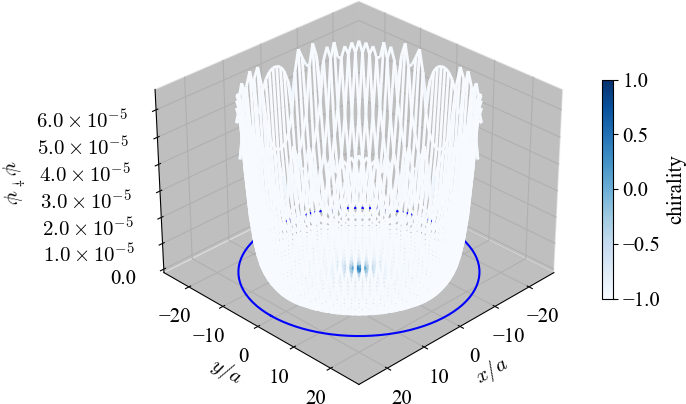}
        \end{minipage}
        \caption{The amplitude of the lowest eigenstate of $D_W^\dagger D_W$ at $r_0=24a$ and $ma=0.35$ (left)
        and that of $D_WD_W^\dagger$ (right)
         The color gradation shows the local chirality.}
            \label{fig:DdaggerDeigenstate_r0=8.0_n=0_m=14perL}
    \end{figure}

Figure \ref{fig:DdaggerDeigenstate_r0=8.0_n=0_m=14perL} presents
the amplitude of the lowest eigenmode at the $z=1/2$ slice
of $D_W^\dagger D_W$ (left panel) and that of $D_WD_W^\dagger$ (right).
The color gradation indicates the local chirality defined by Eq.~(\ref{eq:local chirality}).
As is expected, these eigenmodes are localized at the domain-wall $r=r_0$
and have a uniform distribution of the chirality.

\begin{figure}
\centering
\includegraphics[bb=0 0 461 346]{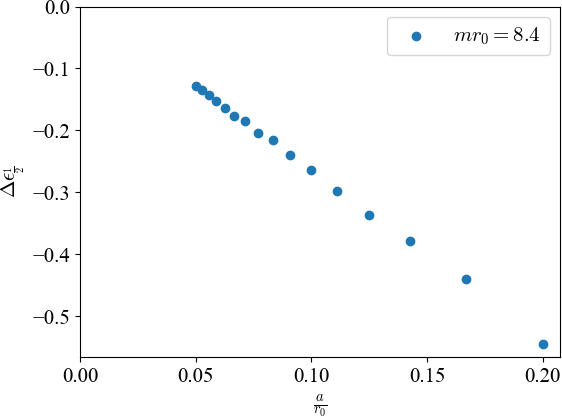}
\caption{The relative deviation of the eigenvalue $E_{1/2}$ as a
function of the lattice spacing $a$ with $mr_0 = 8.4$. 
}
\label{fig:ContinuumLimit}
\end{figure}

In order to estimate the systematic error due to 
the lattice discretization, we vary the lattice spacing
keeping the dimensionless quantity $mr_0=8.4$ fixed.
In Fig.~\ref{fig:ContinuumLimit}, we plot the 
relative ratio of the first eigenvalue,
\begin{equation}
 \Delta \epsilon_j :=\frac{|E_j|-|E_j^{\rm con.}|}{|E_j^{\rm con.}|},
\end{equation}
where $E_j^{\rm con.}$ is the corresponding continuum prediction
obtained by numerically solving Eq.~(\ref{eq:bceq}).
We can see that our lattice values are linearly approaching
the continuum limit with the lattice spacing $a$.

\begin{figure}
\centering
\includegraphics[bb=0 0 461 346]{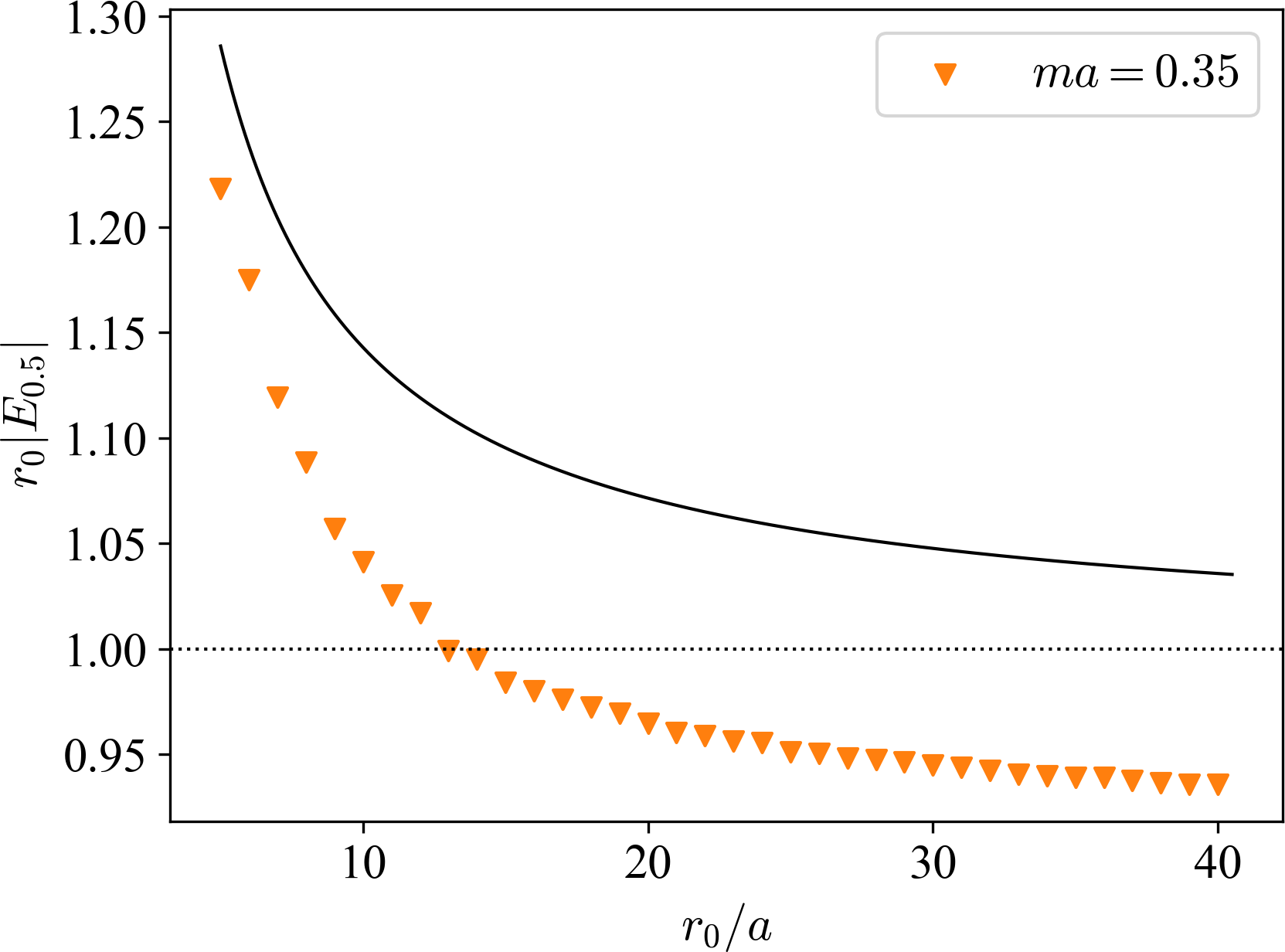}
\caption{The finite volume effect on the eigenvalue of the lowest mode as a function of the radius of the sphere at $ma=0.35$. The solid line represents the continuum prediction $E r_0= 1+ 1/{2mr_0}$.
}
\label{fig:finitevolume}
\end{figure}

In Fig.~\ref{fig:finitevolume}, we plot $r_0|E_{1/2}|$ 
with a fixed lattice spacing $ma=0.35$, as a function of $r_0/a$.
The lattice results in lower triangle symbols are consistent 
with the continuum prediction given by the solid curve, which is
$$
1+\frac{1}{2mr_0}+O(1/(mr_0)^2),
$$ in the large-$mr_0$ expansion.
Note that the finite volume correction is not an exponential but
a power function of $r_0$, since the propagating 
fermion at the edge is massless.


\begin{figure}
    \begin{center}
     \includegraphics[bb=0 0 495 286, height=0.3 \textheight]{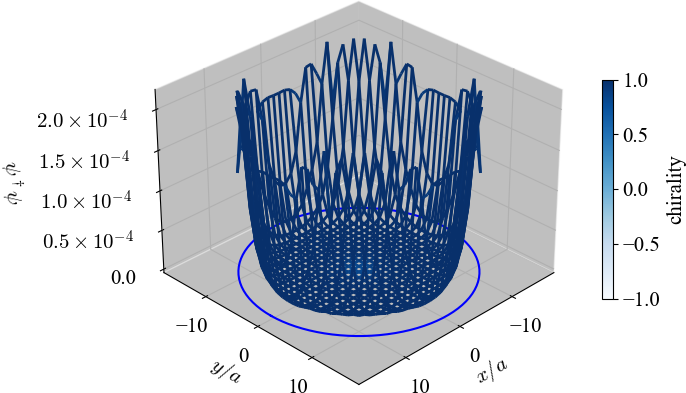}

        \includegraphics[bb=0 0 495 286, height=0.3 \textheight]{DdaggerD/figure/mr0=8.4/eigenstate_r0=24.0a_n=0_ma=0.35000000000000003.png}
        \includegraphics[bb=0 0 494 286, height=0.3 \textheight]{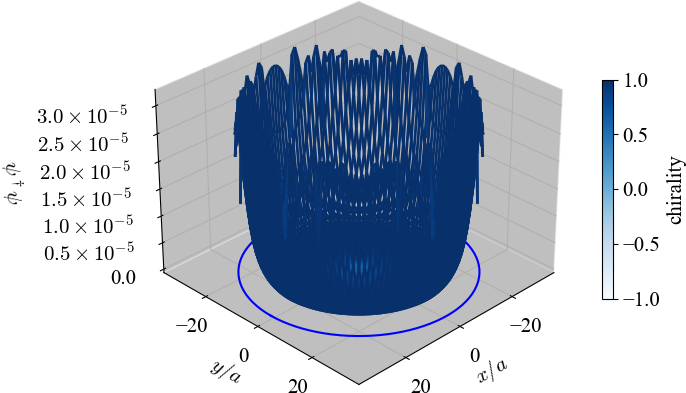}
     \caption{(Top) The amplitude of the lowest eigenstate of $D_W^\dagger D_W$ at $mr_0= 8.4$ and $a/r_0= 1/16$. (Middle) That at $a/r_0= 1/24$. (Bottom) That at $a/r_0= 1/32$.} 
     \label{fig:eigenfunction amplitudes}
    \end{center}
\end{figure}

\begin{figure}
\centering
\includegraphics[bb=0 0 426 286, height=0.3 \textheight]{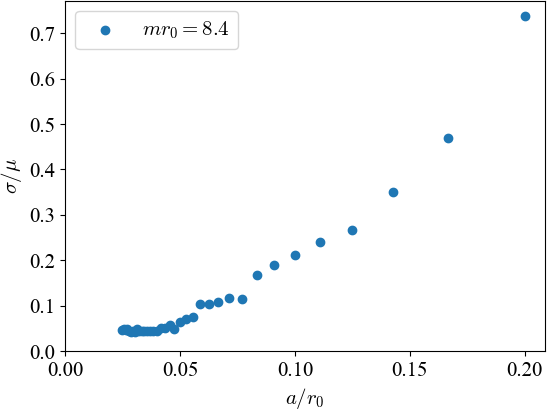}
\caption{The standard deviation $\sigma$ divided by the average $\mu$ of the peaks as a function of the lattice spacing $a$ with $mr_0=8.4$.}
\label{fig:rotational symmetry}
\end{figure}

Finally let us examine the rotational symmetry
in the continuum limit.
The three plots in Fig.~\ref{fig:eigenfunction amplitudes} show the same
eigenfunction with $j=1/2$ with three different lattice spacings
$a/r_0= 1/16,1/24 $ and $1/32$.
The finer the lattice spacing, the better the rotational symmetry looks.
The symmetry can be quantified by taking the standard deviation $\sigma$ of the 
amplitude peaks at the nearest-neighbor points to the domain-wall $r=r_0$ 
normalized by their average $\mu$ as presented in Fig.~\ref{fig:rotational symmetry}.
The result is not monotonic but still shows 
a reasonable convergence to zero in the $a\to 0$ limit.
This supports that our digitized surface 
is a good regularization of the $2D$ sphere.

The above numerical evidence in the free case shows that 
we can formulate a single Weyl fermion 
by this single curved domain-wall fermion on a square lattice.
In the large-$mr_0$ limit, the finite volume effect
is suppressed by $1/mr_0$ and the chirality becomes exact.
But below we will see that the system shows a 
nontrivial behavior when a gauge field is turned on. 

\section{With nontrivial $U(1)$ gauge fields}
\label{sec:LatticeU1}

Next we set nontrivial $U(1)$ gauge link variables,
\begin{align}
    U_j (x)=\exp(i\int_{x}^{x+\hat{j}} A_j(x^\prime) dx^{\prime j}  ) ,
\end{align}
where the vector potential $A_j(x)$ is defined in Eq.~(\ref{eq:Amu}).
The covariant difference operator in the Wilson Dirac operator
acts as
\begin{align}
    (\nabla_j \psi)_{x}= U_j(x) \psi_{x+\hat{j} } - \psi_{x}. 
\end{align}

With these link variables, 
the Dirac string has no physical effect.
In other words, the singularity of the continuum
gauge field is automatically canceled by
the multivaluedness of the $U(1)$ link variables.
To confirm this, let us compare the values of the 
two plaquettes $p_+$ and $p_-$ in the $x$-$y$ plane whose centers are located
at $x_+=(0,0,1/2)$ and $x_-=(0,0,-1/2)$, respectively.
If the Dirac string is physical, the plaquette values
should be asymmetric since the string intersects the plaquette
at $x_-$ but does not for the one at $x_+$.
But this is not true:  by a direct computation, we obtain
the symmetric result:
\begin{align}
 p_\pm &= \exp\left[i\int_{-1/2}^{1/2}dx A_1(x,-1/2,\pm 1/2)+i\int_{-1/2}^{1/2}dy A_2(1/2,y,\pm 1/2)
\right.\nonumber\\& \left.
-i\int_{-1/2}^{1/2}dx A_1(x,1/2,\pm 1/2)-i\int_{-1/2}^{1/2}dy A_2(-1/2,y,\pm 1/2) \right]\nonumber\\
 &= \exp(\pm \pi in/3),
\end{align}
which indicates that the magnetic flux from the monopole
penetrating the two plaquettes at $x_\pm$ is the same.
On this lattice, the total magnetic flux from the monopole
is nonzero, which allows violation of the Bianchi identity
in the continuum limit.

\begin{figure}
    \centering
    \includegraphics[bb=0 0 576 346,width=\textwidth]{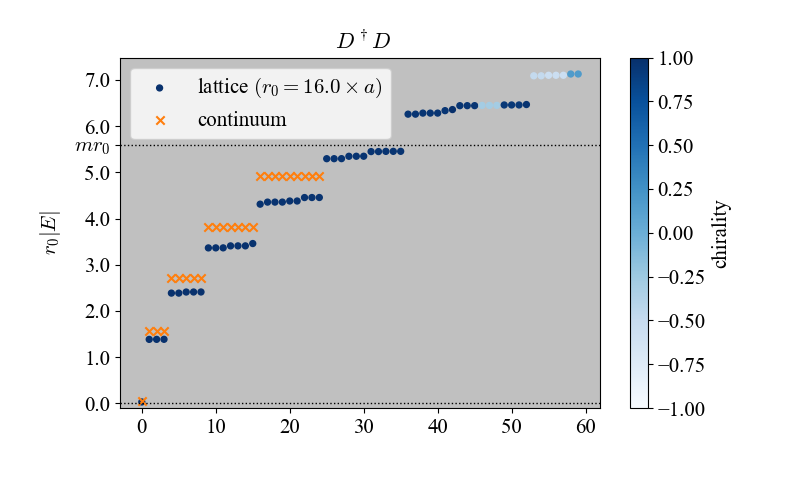}
 \includegraphics[bb=0 0 576 346,width=\textwidth]{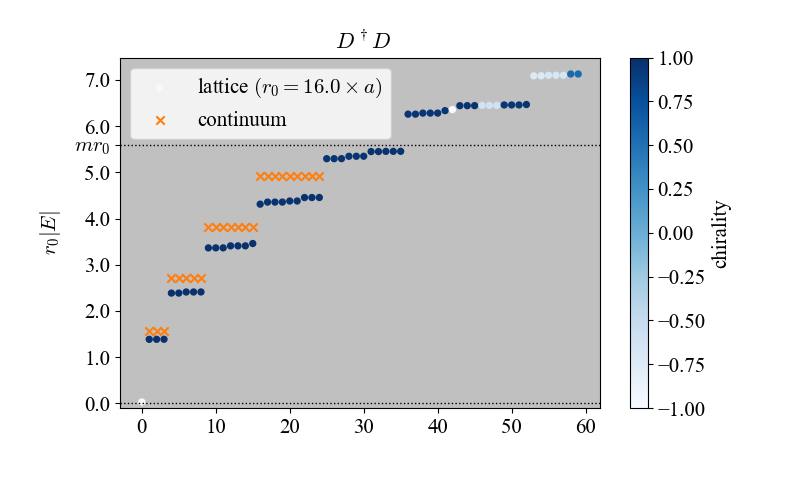}
        \caption{(Top) The eigenvalue spectrum of $D_W^\dagger D_W$ with the monopole charge $n=1$.
Other lattice setup is the same as Fig.~\ref{fig:DdaggerDeigenvalue_r0=8.0_n=0_m=14perL}. (Bottom) 
The same but with $n=-1$.
We can see a near-zero mode, which does not exist in the continuum prediction.}
        \label{fig:eigenvalue_r0=8.0_n=1_m=14perL}
    \end{figure}

In Fig.~\ref{fig:eigenvalue_r0=8.0_n=1_m=14perL}, we plot the eigenvalue
spectrum of $D_W^\dagger D_W$ with the monopole charge $n=1$ (top panel)
and that with $n=-1$ (bottom).
The lattice setup is the same as Fig.~\ref{fig:DdaggerDeigenvalue_r0=8.0_n=0_m=14perL}.
For $n=+1$, the lattice result is consistent with the continuum prediction marked by cross symbols.
The low-lying modes with $r_0|E|<mr_0$ have $\sim +1$ chirality
as the color gradation indicates.
For $n=-1$, however, we find one near-zero mode with the opposite chirality $\sigma_r=-1$,
which does not exist in the continuum analysis. 

\begin{figure}
\centering
\includegraphics[bb=0 0 426 286, width=\textwidth]{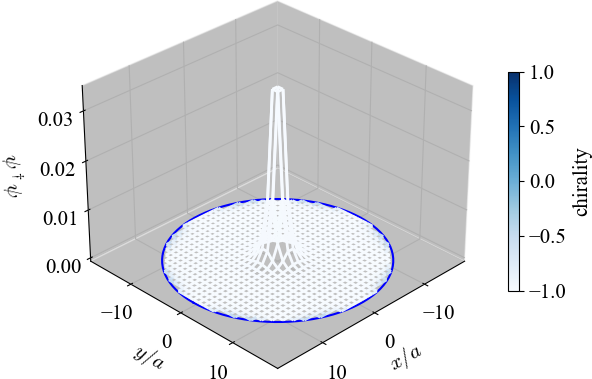}
\caption{The amplitude of the zero mode of $D_W^\dagger D_W$ at $r_0 =16a,~ma=0.35$ and $n=-1$. }
\label{fig:amplitude center-localized mode}
\end{figure}

The amplitude of the oppositely chiral near-zero mode at the $z=1/2$ slice 
is plotted in Fig.~\ref{fig:amplitude center-localized mode}.
This mode is not located at the domain-wall but localized at the center
where the monopole sits.

What is the origin of the new center-localized zero mode with the opposite chirality?
In fact, in our previous work \cite{Aoki:2023lqp}, 
we observed a similar phenomenon in the four-spinor system
where a zero Dirac spinor mode appears in the vicinity of the monopole.
Our microscopic analysis revealed that the additive mass renormalization
from the Wilson term becomes strong enough at the singularity of the gauge field
to flip the sign of the effective mass, which 
creates another small but finite domain-wall near the origin.
This dynamically created domain-wall captures one edge-localized zero mode (of the electron),
which explains why the monopole becomes dyon in topological insulators \cite{Witten:1979Dyons}.

\begin{figure}
\centering
\includegraphics[bb=0 0 426 286,width=0.9\textwidth]{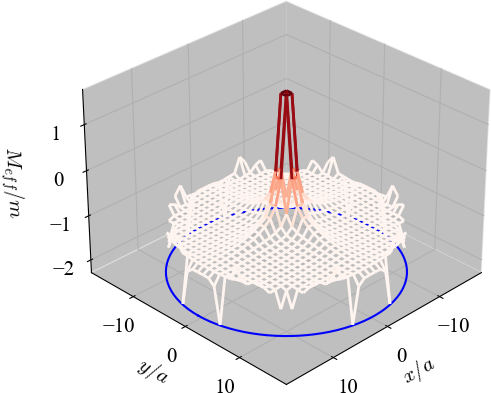}
\includegraphics[bb=0 0 426 286,width=0.9\textwidth]{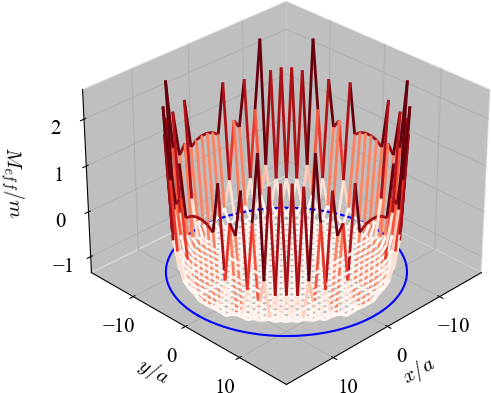}
\caption{(Top) The effective mass of the zero mode at $n=-1$. (Bottom) That of the lowest mode at $n=0$.
The red-white color gradation expresses the sign of the effective mass.}
\label{fig:effectiveMass_r0=8.0_n=1_m=14perL}
\end{figure}

In the top panel of Fig.~\ref{fig:effectiveMass_r0=8.0_n=1_m=14perL}, we present the 
distribution in the $z=1/2$ slice of the effective mass defined by
\begin{align}
    M_{\rm eff}(x)= \frac{\phi_0^\dagger(x) \qty[ \sum_{i=1}^3 \frac{1}{2a} \nabla_i \nabla_i^\dagger -m] \phi_0(x)}{\phi_0^\dagger(x) \phi_0(x)},
\end{align}
where $\phi_0$ denotes the center-localized zero mode with $n=-1$.
We can see that an island of the positive mass region whose
edge is another domain-wall is created. 
For comparison, we also plot the same quantity but with 
the lowest eigenmode with $n=0$ in the bottom panel of 
Fig.~\ref{fig:effectiveMass_r0=8.0_n=1_m=14perL},
where the mass is negative everywhere except for the edge $r=r_0$.
Thus, we can identify the center-localized zero mode
as the edge mode of the new domain-wall near the monopole.

\begin{figure}
    \centering
    \includegraphics[bb=0 0 576 346,width=\textwidth]{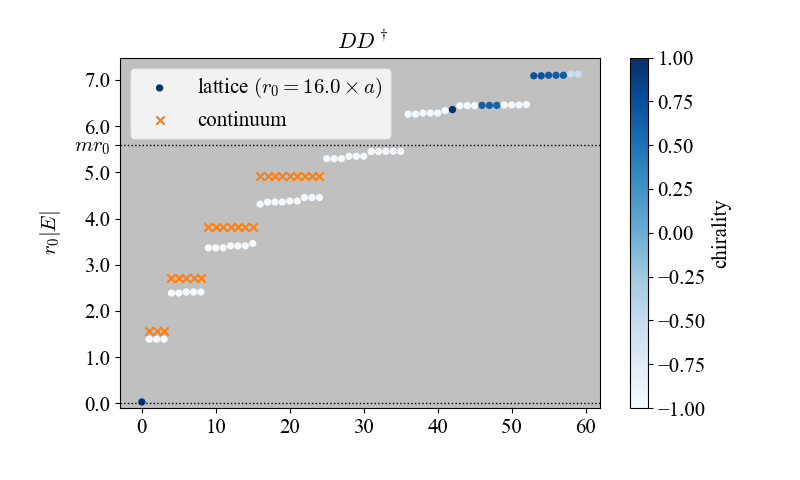}
 \includegraphics[bb=0 0 576 346,width=\textwidth]{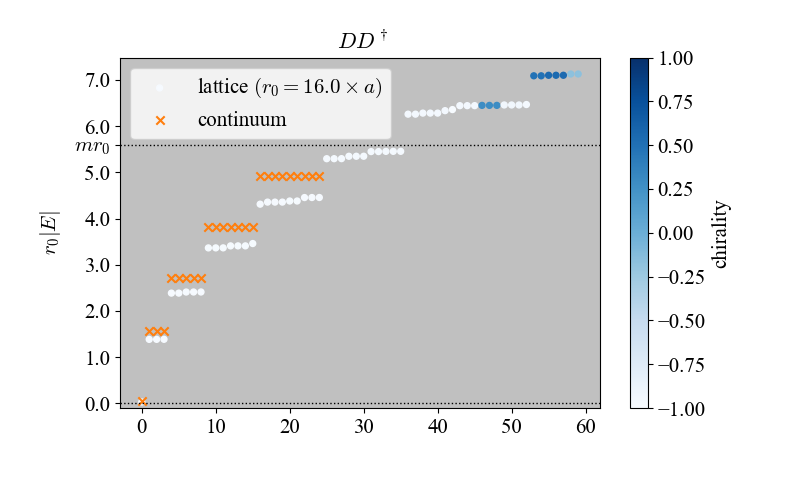}
        \caption{(Top) The eigenvalue spectrum of $D_W D_W^\dagger$ with the monopole charge $n=1$.
Otherwise, the lattice setup is the same as Fig.~\ref{fig:DdaggerDeigenvalue_r0=8.0_n=0_m=14perL}. (Bottom) 
The same but with $n=-1$.}
        \label{fig:eigen-DDdag}
    \end{figure}

Figure~\ref{fig:eigen-DDdag} shows the eigenvalue distribution 
of $D_WD_W^\dagger$. 
A similar center-localized mode with positive chirality, which is opposite to
that localized at $r=r_0$ appears when $n=+1$.

The appearance of another domain-wall and 
the opposite chiral zero mode on it
may indicate that the low-energy theory is 
not a simple chiral theory with 
a single Weyl fermion on a single domain-wall
but a nontrivial vectorlike theory on
the two domain-walls.

\section{How to avoid opposite chiral modes}
\label{sec:discussion}

In this section we show in continuum analysis that
the center-localized mode is stable against perturbation
of the background gauge and gravity field
since it is topologically protected.
It is, therefore, crucial for constructing a chiral gauge theory 
in this setup, to find a formulation 
which eliminates such oppositely chiral zero modes.

In order to make the problem explicit, 
let us consider a Hermitian operator in continuum
\begin{equation}
 \hat{D} :=\left(\begin{array}{cc}
  0& D^\dagger \\
D & 0	   \end{array}\right),
\end{equation}
where the domain-wall Dirac operator $D$ is defined as per Eq.~(\ref{eq:D})
but with two domain-walls:
\begin{align}
\label{eq:massterm2}
  m(x)=\left\{
  \begin{array}{cc}
  -m & \text{for $r_1 \le |x|\le r_0$}\\
  +M & \text{otherwise}
  \end{array}\right.,
\end{align}
where we assume that $r_1$ is put in the vicinity of the monopole
at the inverse of the cutoff scale.

In the $M\to +\infty$ limit, the edge-localized states
are the eigenmode of
\begin{equation}
 \gamma := \sigma_3 \otimes \sigma_r. 
\end{equation}
The positive $\gamma$ modes are located at the outer domain-wall at $r=r_0$,
whereas the negative $\gamma$ modes are at $r=r_1$. 
At the outer domain-wall, the Dirac operator acts as
the $2D$ massless operator Eq.~(\ref{eq:2DD})
and the same is true for $r=r_1$ but with different scale:
\begin{equation}
 \bar{D}^{S^2} = \frac{1}{r_1}\left[\sigma^i \left(L_i+n\frac{x_i}{2r}\right) +1\right].
\end{equation}
It is not difficult to see that the two massless Dirac operator
$D^{S^2}$ at $r=r_0$ and $\bar{D}^{S^2}$ at $r=r_1$ share the same index:
\begin{align}
 {\rm Ind}D^{S^2} ={\rm Ind}\bar{D}^{S^2} &= \frac{1}{4\pi}\int_{S^2}d^2x \epsilon^{\mu\nu}F_{\mu\nu}=n.
\end{align}
Namely, the existence of the zero modes at the two spherical domain-walls
is topologically guaranteed by the AS index on them.
For $n> 0$, the two zero modes with $\sigma_r={\rm sgn}(n)$ 
appear for $D$ on the $r=r_0$ domain-wall and another for
 $D^\dagger$ is localized at $r=r_1$.
For $n<0$, there are also two zero modes but the roles of the $r_0$ and $r_1$ are flipped.

The above equivalence 
${\rm Ind}D^{S^2} ={\rm Ind}\bar{D}^{S^2}$ is a consequence
of cobordism invariance of the AS index.
Since the two spheres at $r=r_0$ and $r=r_1$ share
the same bulk region in $r_1<r<r_0$,
they are cobordant and must have the same index.

Therefore, the opposite chiral zero modes are as stable as
those localized at the edge $r=r_0$
against any perturbative deformation of the configuration.
Their existence in this continuum analysis looks like a serious obstacle in
constructing a chiral gauge theory.
The situation is apparently the same as
the flat two domain-wall system where
it is difficult to decouple the unwanted opposite chiral modes
from the theory.
Some times ago a promising proposal was made, to use the Yang-Mills gradient flow
to turn off the gauge interaction for one of the chiral modes in the two domain-wall fermion system \cite{Grabowska:2015qpk,Grabowska:2016bis}.
However it was pointed out in Ref.~\cite{Okumura:2016dsr}
(see also related works \cite{Hamada:2017tny, Makino:2016auf,Makino:2017pbq, Ago:2019yzj})
that a topologically nontrivial gauge background
cannot eliminate the zero modes that originated from the AS index.
This is also a consequence of the cobordism invariance of the index. 

Note, however, that the situation on the lattice
is not exactly the same as the continuum analysis above.
First, the inner domain-wall is not fixed but can appear only with 
singular gauge field configuration.
Second, the center-localized mode is 
effectively a zero-dimensional object, having no $\theta$ and $\phi$ dependence
on the small sphere, which is
isolated from the bulk modes.
Taking these differences from the flat two-domain-wall case into account,
we would like to propose two ideas which may be useful to avoid
appearance of the opposite chiral modes.

One is to impose the admissibility condition \cite{Luscher:1981zq} 
or a smoothness condition for every plaquette on the lattice.
Under this condition, the only admissible link gauge fields,
satisfying
\begin{align}
||1-P_{ij}(x)|| < \epsilon,
\end{align}
with some small real number $\epsilon$ 
are taken to construct the theory.
Then, singular configurations are
not allowed in the theory.
In the continuum limit, the condition is equivalent to
imposing the Bianchi identity.

Under the admissibility condition, we cannot put
a monopole on the lattice.
If $\epsilon$ is small enough, we will be able to 
limit the additive mass renormalization through the Wilson term
and avoid dynamical creation of the domain-walls.
However, this would restrict the surface theory 
to be fixed in a topologically trivial sector 
where the instanton number is forced to be zero.
By enlarging the gauge symmetry, which is a well-known prescription
by 't~Hooft and Polyakov \cite{t_Hooft1974-yf,Polyakov1974-up}, 
this fixing topology problem may be circumvented but
the discussion is beyond the scope of this work.

The second proposal is to resort to 
the so-called symmetric mass generation.
Recently many studies \cite{
Wen:2013ppa,Wang:2013yta,Kikukawa:2017ngf,Kikukawa:2017gvk,Wang:2018jkc,you2018symmetric,Tong:2021phe,Zeng:2022grc,Wang:2022ucy,Lu:2023emm,Guo:2023rnz,Lu:2023cev,Liu:2023msa,Golterman:2023zqf}
have discussed possibility of gaping out the edge-localized modes
for special combinations of 
Weyl fermions by nontrivial interactions such as 
four-Fermi vertices, without breaking the chiral symmetry.
This is possible only when their both perturbative 
and nonperturbative gauge anomalies
are canceled.
Since the unwanted center-localized modes
in our system are essentially zero-dimensional,
the analysis may be a lot easier than for those in general dimensions.
Once the center-localized zero modes were gaped out,
they would have an eigenvalue which scales as $1/r_1$,
and we will be able to show that they are decoupled from
the low-energy dynamics and the total system is essentially chiral.

\section{Summary and discussion}
\label{sec:summary}

In this work, we have investigated a Shamir lattice 
domain-wall fermion system with a single curved surface.
We have embedded a $2D$ spherical domain-wall
into a $3D$ flat square lattice ignoring
fermion hopping to the outside of the domain-wall.
Solving the Dirac equation of
the $3D$ negatively massive Wilson fermion
in this system, we have examined the existence of massless
chiral edge-localized modes and the gravity
they feel through the induced spin connections.

In the free fermion theory with trivial gauge link variables,
we have verified the existence of the edge-localized 
modes on the surface.
Moreover, we have shown that 
these edge modes are almost chiral and massless.
We have also identified the effect of the induced gravity
as a gap in the Dirac eigenvalue spectrum.
The continuum extrapolation and large volume limit
are well under control, including recovery of the
rotational symmetry.
Our single domain-wall fermion system thus
describes a single Weyl fermion in the low-energy limit.

With nontrivial $U(1)$ gauge link variables, however,
it has turned out that the system shows a dramatic change.
With the magnetic monopole-like configuration, we
have observed that a center-localized zero mode appears
in the vicinity of the monopole, having the opposite
chirality to that of the edge modes.
This was not expected in the continuum analysis, 
under a condition where the eigenfunctions are smooth everywhere.
We have found that the singularity of the gauge field
creates another small but finite domain-wall via
additive mass renormalization near the
origin, which makes the massless zero mode localized at the wall.

We have discussed the stability of the dynamically created domain-wall and
the center-localized zero mode on it
and shown that they are topologically protected
by the AS index theorem on the sphere.
We have then made two proposals to avoid the appearance of these
opposite chiral zero modes.
One is to impose the admissibility condition, and another
is the use of symmetric mass generation.
The smallness of the domain-wall radius may help in 
the analysis of the center-localized mode as it is essentially 
a zero-dimensional object, which is greatly isolated from the massive bulk modes.

We thank S.~Aoki, M.~Furuta, S.~Iso, D.B.~Kaplan, Y.~Kikukawa, M.~Koshino, Y.~Matsuki, S.~Matsuo, T.~Onogi, S.~Yamaguchi, M.~Yamashita and R.~Yokokura for useful discussions.
In particular, we thank Y.~Kikukawa for his instruction to 
technical details on the symmetric mass generation.
The work of SA was supported by JSPS KAKENHI Grant Number JP23KJ1459. The work of HF and NK
was supported by JSPS KAKENHI Grant Number JP22H01219.




\bibliographystyle{ptephy}
\bibliography{ref_submit}

\end{document}